\documentclass{PoS}

\title{%
\vspace*{-1cm}
\begin{minipage}{\textwidth}
\begin{flushright}
\texttt{\footnotesize
PoS(Confinement X)111\\%
}
\end{flushright}
\end{minipage}\\[15pt]
Determination of SU(2) ChPT LECs\\ from 2+1 flavor staggered lattice simulations}

\ShortTitle{Determination of SU(2) ChPT LECs from 2+1 flavor staggered lattice simulations}

\author{\speaker{Enno~E.~Scholz}$^a$, Szabolcs~Bors\'anyi$^b$, Stephan~D\"urr$^{b,c}$, Zolt\'an~Fodor$^{b,c,d}$, Stefan~Krieg$^{b,c}$, Andreas~Sch\"afer$^a$, and Kalman~K.~Szabo$^b$\\
\llap{$^a$}Universit\"at Regensburg, Universit\"atsstr.~31, D-93053 Regensburg, Germany\\
\llap{$^b$}Bergische Universit\"at Wuppertal, Gau\ss{}str.~20, D-42119 Wuppertal, Germany\\
\llap{$^c$}J\"ulich Supercomputing Centre, Forschungszentrum J\"ulich, D-52425 J\"ulich, Germany\\
\llap{$^d$}Institute for Theoretical Physics, E\"otv\"os University, H-1117 Budapest, Hungary\\
E-mail: \email{enno.scholz(AT)physik.uni-regensburg.de}%
, \email{durr(AT)itp.unibe.ch}
}

\abstract{%
By fitting pion masses and decay constants from 2+1 flavor staggered lattice simulations to the predictions of NLO and NNLO SU(2) chiral perturbation theory we determine the low-energy constants $\bar{\ell}_3$ and $\bar{\ell}_4$. The lattice ensembles were generated by the Wuppertal-Budapest collaboration and cover pion masses in the range of 135 to 435 MeV and lattice scales between 0.7 and 2.0 GeV. By choosing a suitable scaling trajectory, we were able to demonstrate that precise and stable results for the LECs can be obtained from continuum ChPT to NLO. The pion masses available in this work also allow us to study the applicability of using ChPT to extrapolate from higher masses to the physical pion mass.}

\FullConference{Xth Quark Confinement and the Hadron Spectrum,\\
		 October 8-12, 2012\\
		 TUM Campus Garching, Munich, Germany}

\newlength{\closercaption}
\newlength{\afterTable}
\newlength{\afterFigure}
\newlength{\closersection} 
\begin{document}

\section{Introduction}
\label{sec:intro}
\vspace*{\closersection}

Chiral perturbation theory (ChPT) \cite{Gasser:1983yg,Gasser:1984gg} is a widely used tool in many phenomenological applications and also helpful to guide an extrapolation to lighter quark masses in lattice-QCD simulations. Here we will report on a determination of the NLO low-energy constants (LECs) $\bar{\ell}_3$ and $\bar{\ell}_4$ which appear in the light quark mass dependence of the pseudo-scalar meson masses and decay constants in SU(2) ChPT.

\begin{figure}[b!]
\begin{center}
\begin{minipage}{.45\textwidth}
\includegraphics*[width=\textwidth]{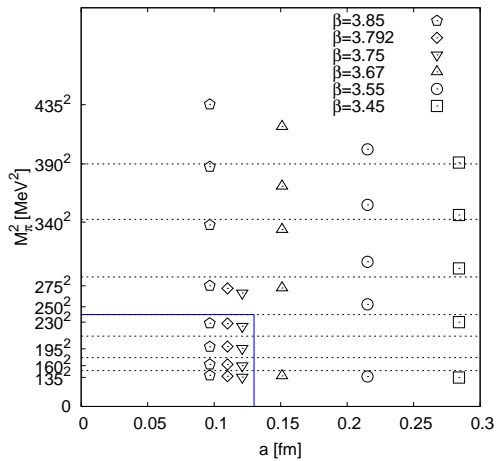}
\end{minipage}%
\hspace*{.1\textwidth}
\begin{minipage}{.4\textwidth}
\caption{Simulated pion masses squared $M_\pi^2$ vs.\ lattice spacing $a$ at six different gauge couplings $\beta$. {\it Horizontal dashed lines} indicate cuts on the mass range in our ChPT fits. The lower left corner marked by {\it blue solid lines} is our final preferred fit range. See text for details.}
\label{fig:landscape}
\end{minipage}
\end{center}
\end{figure}

We analyze configurations generated by the Wuppertal-Budapest Collaboration \cite{Aoki:2006we,Aoki:2006br,Aoki:2005vt,Aoki:2009sc,Borsanyi:2010bp,Borsanyi:2010cj} using the Symanzik glue and 2-fold stout-smeared staggered fermion action for a 2+1 flavor QCD-simulation. The mass of the single flavor has been kept at the value of the physical strange quark mass, whereas the two degenerate lighter quark masses have been varied such that light meson masses in the range of 135 to 435 MeV were simulated. The simulations were performed at six different gauge couplings $\beta$, resulting in lattice scales between 0.7 and 2.0 GeV (see next section for details on how the scale has been determined). Figure \ref{fig:landscape} shows a landscape plot of our simulated pion masses squared versus the lattice spacing. Details about the simulated gauge couplings, lattice volumes, and tuning of the input quark mass values are reported in our publication \cite{Borsanyi:2012zv}.

The 2-fold stout-smeared version of the staggered quark action has been proven to be advantageous \cite{Borsanyi:2010bp} in reducing the inevitable taste-breaking of staggered fermion formulations. Therefore, in this work we only consider the pseudo-scalar mesons with taste matrix $\gamma_5$ when measuring meson masses or decay constants. Again, for details of the computation of these quantities we refer the reader to \cite{Borsanyi:2012zv}.

\vspace{\afterFigure}

\section{Scale setting and physical quark masses}
\label{sec:scale_mud}
\vspace*{\closersection}

\begin{figure}[t!]
\begin{center}
\includegraphics*[width=.45\textwidth]{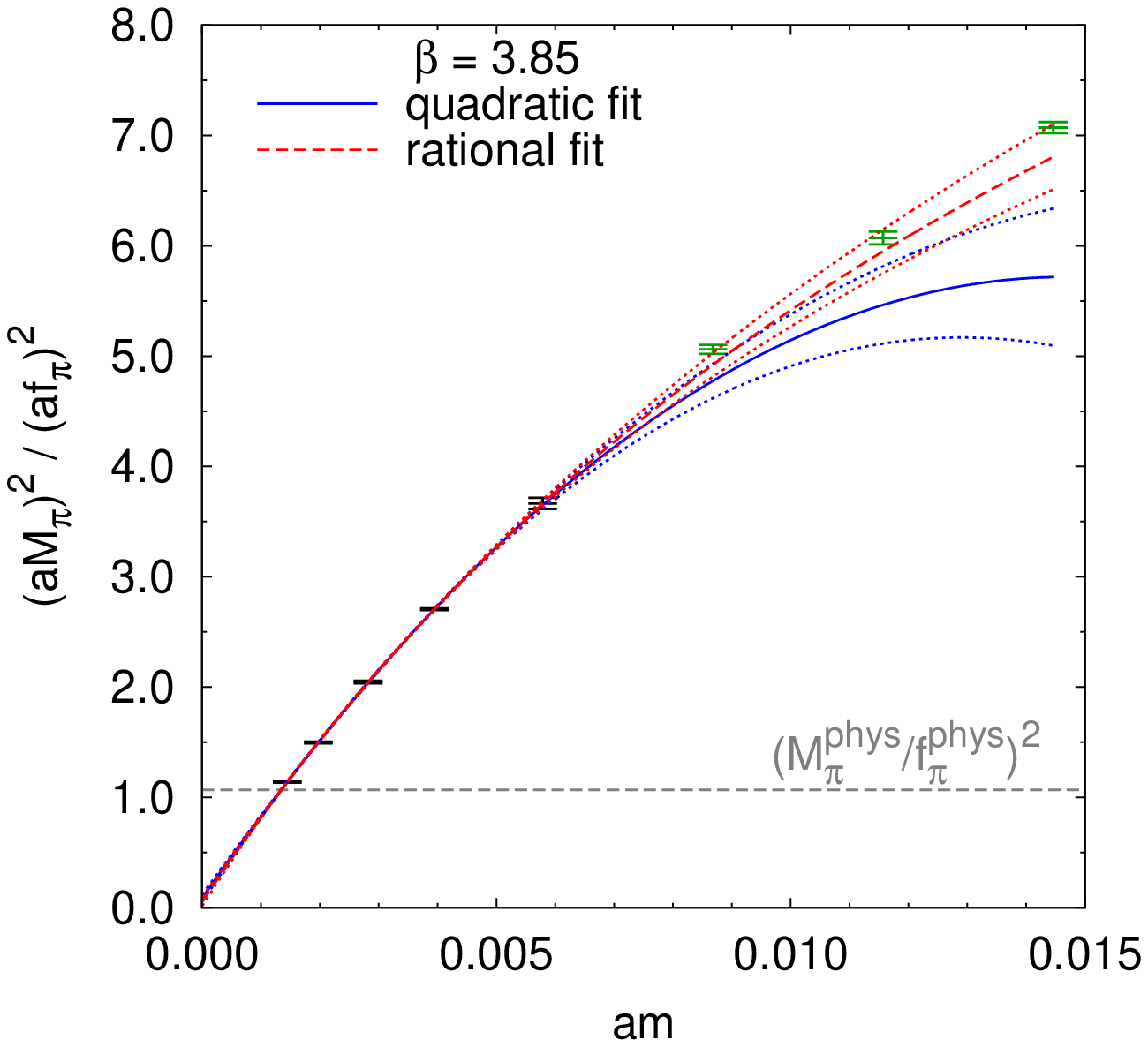}%
\includegraphics*[width=.45\textwidth]{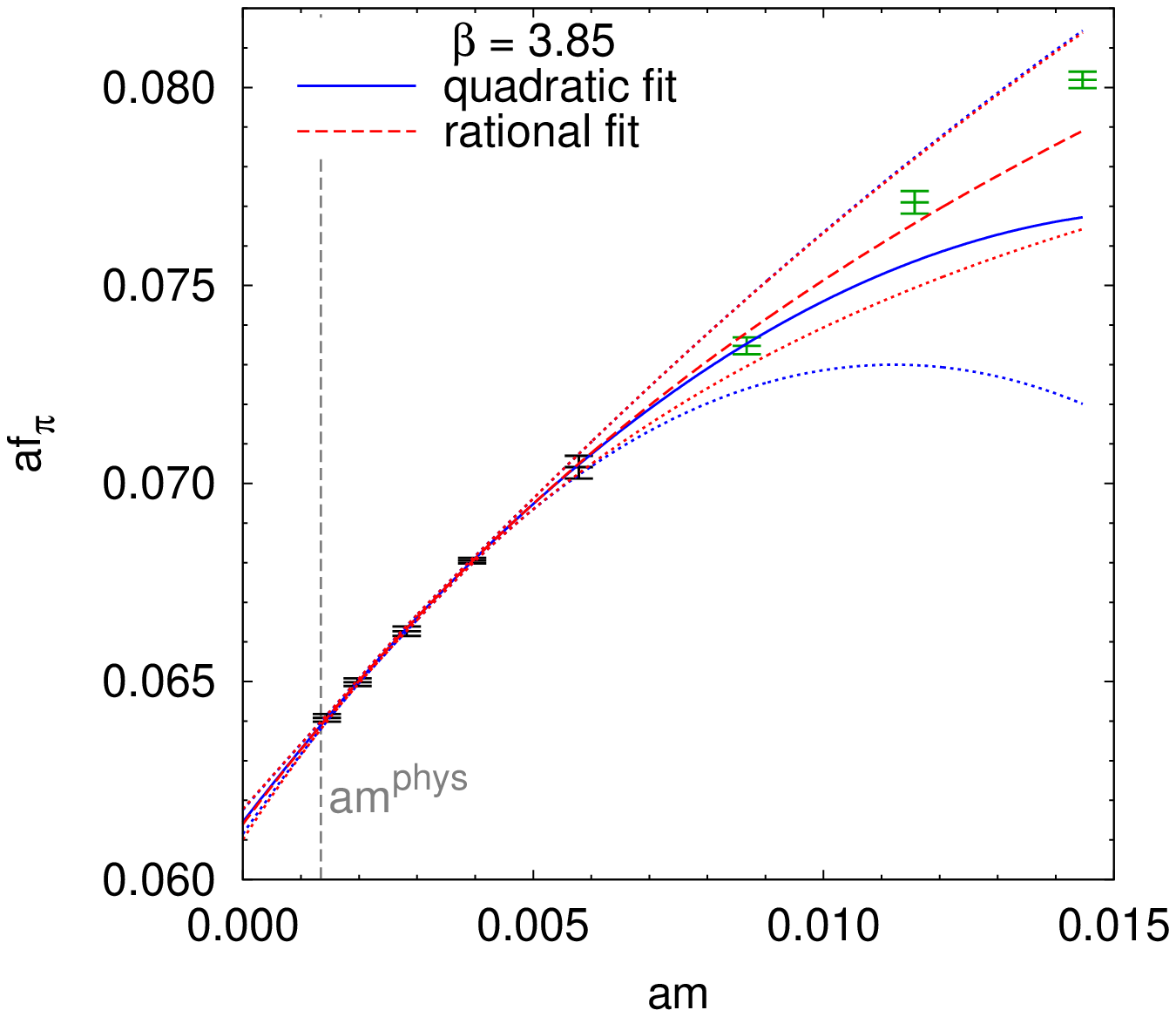}%
\end{center}
\vspace*{\closercaption}
\caption{{\it Left panel:} ratio $(aM_{\pi})^2/(af_{\pi})^2$ extrapolated to $(M_\pi^{\rm phys}/f_\pi^{\rm phys})^{2}=1.06846$ to obtain $am^{\rm phys}$, {\it right panel:} $af_{\pi}$ extrapolated to $am^{\rm phys}$ to obtain $1/a$; both at $\beta=3.85$. Points marked by {\it black symbols} are included in the fit, while those marked by {\it green symbols} are excluded.}
\label{fig:set_mud_scale}
\end{figure}

To set the scale at each simulated gauge coupling $\beta$ and identify the physical point, i.e.\ the average up/down quark mass $m^{\rm phys}=(m_{\rm u}+m_{\rm d})/2$ corresponding to a pion in the isospin limit with an estimated mass of $M_\pi^{\rm phys}=134.8\,{\rm MeV}$ \cite{Colangelo:2010et}, we use a two-step procedure. First, we extrapolate the ratio $(aM_{\pi})^2/(af_{\pi})^2$ of the measured squared meson masses and decay constants to its physical value $(134.8\,{\rm MeV} / 130.41\,{\rm MeV})^2 = 1.06846$, where we also used the PDG-value $f_\pi^{\rm phys}=130.41\,{\rm MeV}$ \cite{Nakamura:2010zzi}. In that way $am^{\rm phys}$ is obtained. In the second step, we extrapolate $af_{\pi}$ to this quark mass value and obtain the lattice scale with the help of the PDG-value $f_\pi^{\rm phys}$. For the extrapolation we used two different ans\"atze: a quadratic and a rational (linear in numerator and denominator) fit form. An example of these extrapolations is shown for the ensembles at $\beta=3.85$ in Fig.~\ref{fig:set_mud_scale}. There, only the five lightest points were used in the fits, which is a typical choice for all other ensembles. We stress that here, like in the chiral fits to be discussed below, the data has been corrected for finite volume effects beforehand, by means of using the two- and three-loop resummed formulae of \cite{Colangelo:2005gd} for the pion decay constants and masses, respectively. Our spatial lattice volumes $L^3$ are in the range $(4.3\,{\rm fm})^3$ -- $(6.8\,{\rm fm})^3$ with a minimal $M_{\pi}L\approx 3.3$. This ensures that we only observe small finite volume corrections. In case of the pion mass the correction factors vary between 0.1 and 2.7 per-mille and in case of the decay constant between 0.2 and 7.5 per-mille.

By fixing $1/a$ and $am^{\rm phys}$ in the way described above, the meson masses and decay constants show no discretization effects at all directly at the physical point and we can assume those effects to be small (since of higher order in the quark masses and/or lattice spacing) in the vicinity of the physical point, i.e.\ in the mass range covered by our fits. Such discretization effects, of course, are present in other observables, which are not considered in this work.

\vspace*{\afterFigure}

\section{Fits to NLO SU(2) ChPT}
\label{subsec:fits.nlo}
\vspace*{\closersection}

The quark mass dependence of the finite-volume corrected data for the meson masses and decay constants is fitted simultaneously at different $\beta$-values using the NLO-SU(2) ChPT formulae
\begin{eqnarray*}
M_{\pi}^2 &=& \left(\frac{1}{a}\right)^2 (aM_{\pi})^2 \;=\; \chi\,\left[1\,+\,\frac{\chi}{16\pi^2 f^2}\log\frac{\chi}{\Lambda_3^2}\right]\,, \\
f_{\pi} &=& \left(\frac1{a}\right) (af_{\pi}) \;=\; f\left[1\,-\,\frac{\chi}{8\pi^2f^2}\log\frac{\chi}{\Lambda_4^2}\right]\,,\;\;\;
\chi \;=\; 2B\,m \;=\; (2Bm^{\rm phys})\,\frac{am}{am^{\rm phys}}\,,
\end{eqnarray*}
where we made use of the already determined $1/a$ and $am^{\rm phys}$ to scale the quark masses and the meson masses and decay constants measured in lattice units. This fit has four free parameters: two NLO low-energy scales $\Lambda_3$, $\Lambda_4$ (related to the LECs $\bar{\ell}_i=\log[\Lambda_i^2/(M_\pi^{\rm phys})^2]$), the decay constant in the SU(2) chiral limit $f$ and the renormalization scheme-independent combination $\chi^{\rm phys}\,=\,(2Bm^{\rm phys})$ of the LO low-energy constant $B$ and the physical quark mass $m^{\rm phys}$. Since we used the physical pion mass and decay constant to set the scale at each set of ensembles with a common $\beta$, and furthermore each set contains at least one data point in close vicinity to the physical point, our ansatz should reproduce the physical point. Therefore, we also used a parameter-reduced chiral fit, where this constraint has been implemented and only two free fit parameters remain, which we chose to be $\chi^{\rm phys}$ and $f$. Our exact implementation of the parameter-reduced fit formulae is reported in \cite{Borsanyi:2012zv}. 

We would like to point out that the chiral fit formulae do not include any taste breaking effects, i.e., we did not use staggered ChPT. This seems justified to us, since we are only considering $\gamma_5$-taste mesons as mentioned above and use these to define our scaling trajectory at the physical point. In other words, since the meson mass and decay constant at the physical point were used to set the quark masses and lattice scales, no discretization or taste breaking effects are present in the ChPT formulae for $M_{\pi}^2$ and $f_{\pi}$ as discussed above. Furthermore, taste breaking effects are reduced anyway by the choice of the fermion action as mentioned above.

\begin{figure}[t!]
\begin{center}
\includegraphics[width=.45\textwidth]{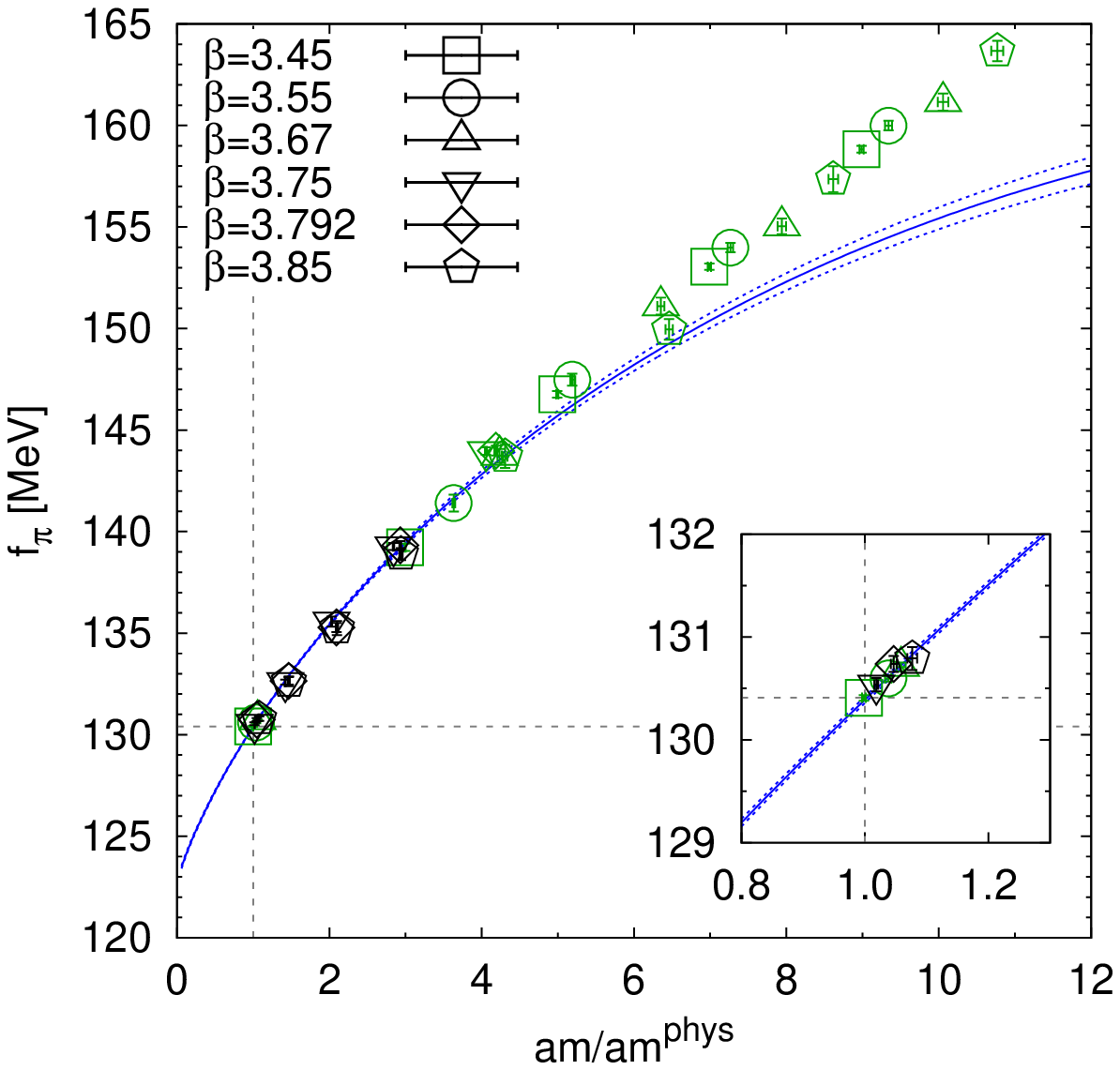}%
\includegraphics[width=.45\textwidth]{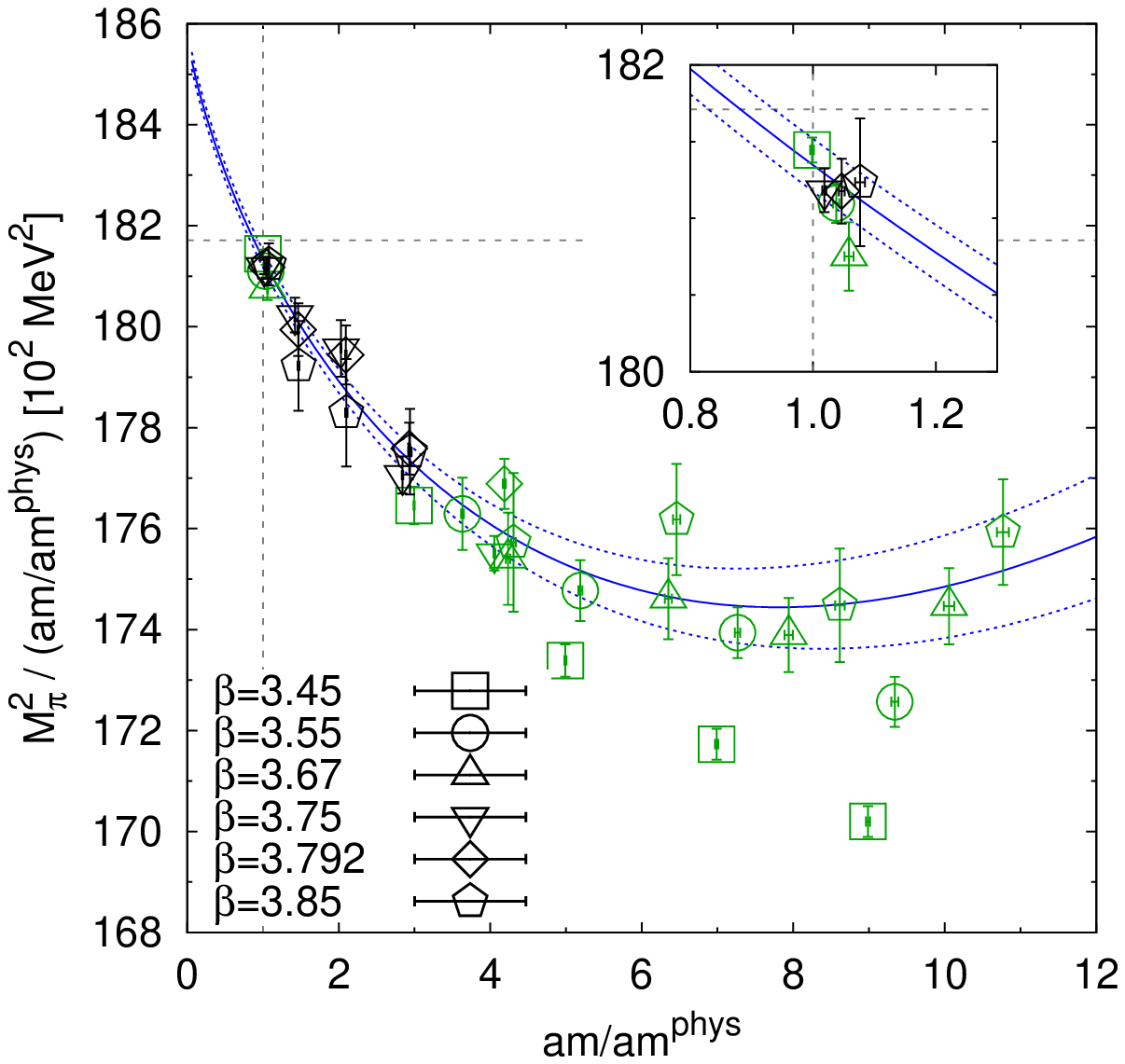}
\end{center}
\vspace*{\closercaption}
\caption{Combined global unconstrained NLO fit of the decay constants ({\it left panel}) and squared meson masses divided by the quark mass ratio ({\it right panel}) to ensembles with $1/a\geq 1.6\,{\rm GeV}$ and $135\,{\rm MeV}\leq M_\pi\leq 240\,{\rm MeV}$. Points included in the fit range are marked by {\it black symbols}, points excluded by {\it green symbols}.}
\label{fig:fitNLO_135_240_fine}
\end{figure}
\vspace*{\afterFigure}

\begin{figure}[t!]
\begin{center}
\includegraphics[width=.4\textwidth]{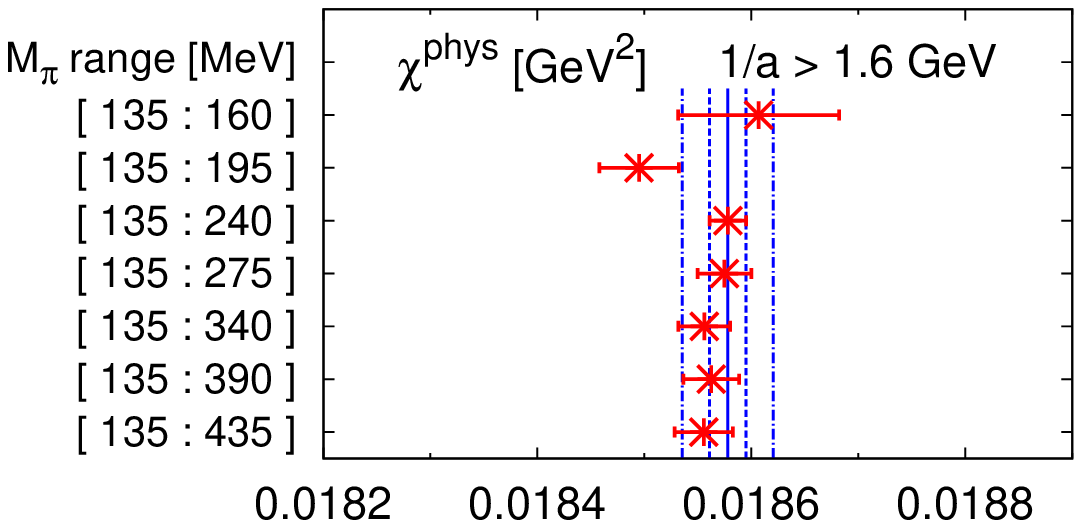}%
\includegraphics[width=.4\textwidth]{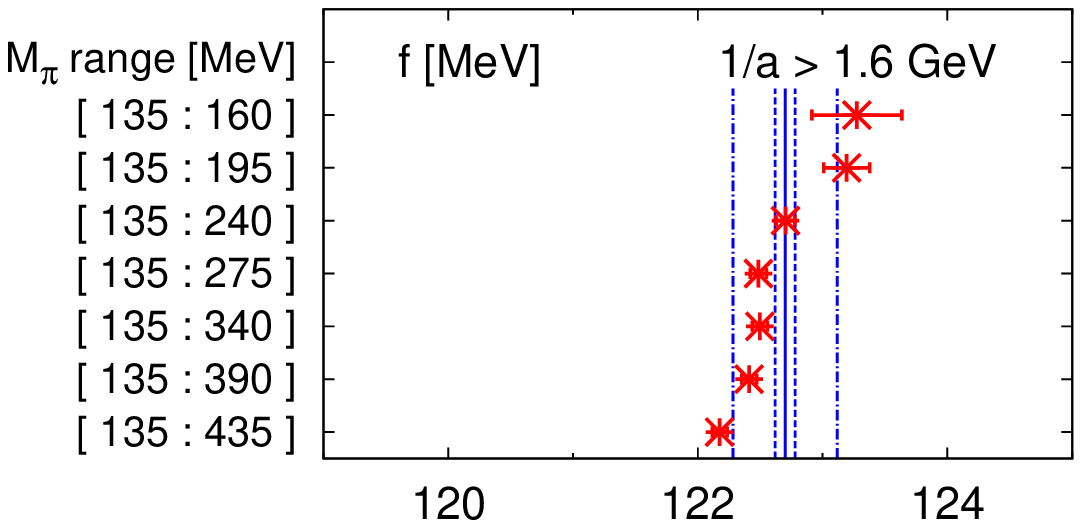}\\
\includegraphics[width=.4\textwidth]{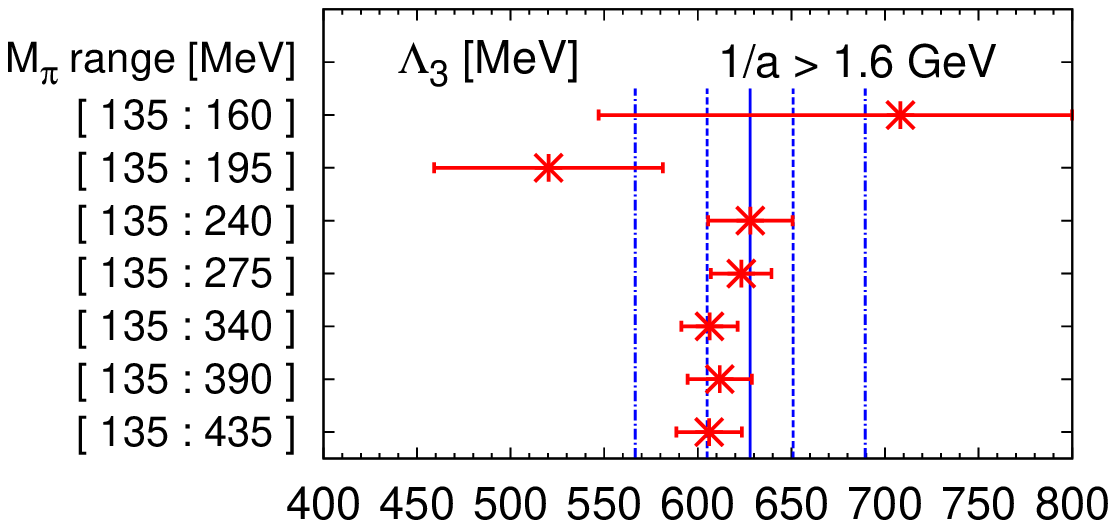}%
\includegraphics[width=.4\textwidth]{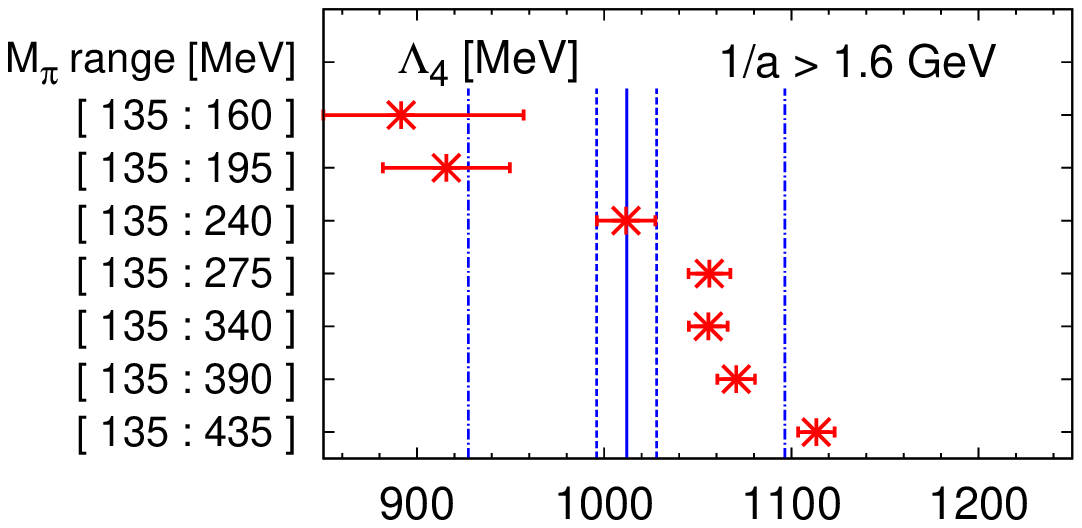}\\
\begin{minipage}{.4\textwidth}
\includegraphics[width=\textwidth]{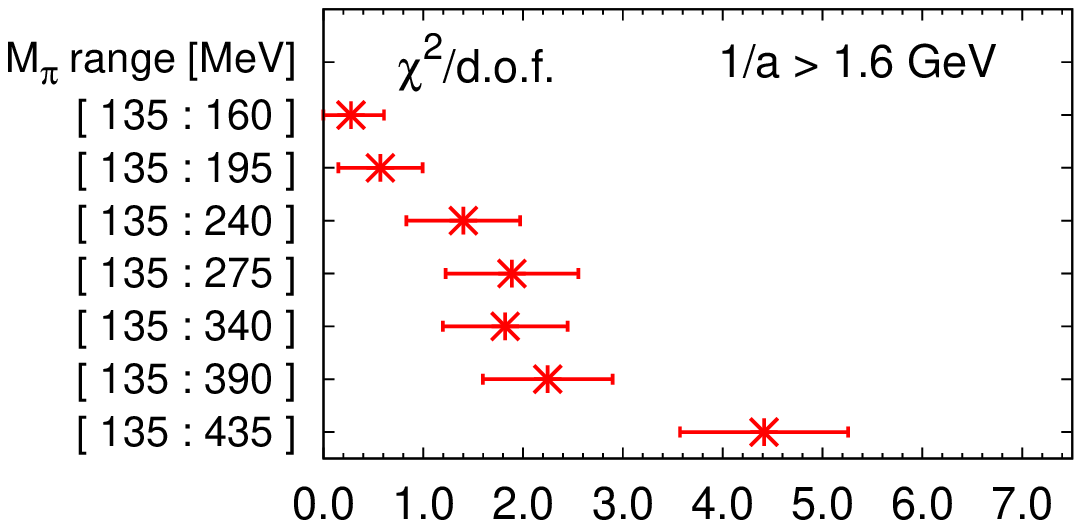}
\end{minipage}%
\begin{minipage}{.55\textwidth}
\caption{Dependence of the fitted parameters and (uncorrelated) $\chi^2/{\rm d.o.f.}$ on the mass range of fits only including ensembles with $1/a\geq 1.6\,{\rm GeV}$. The {\it solid}, {\it dashed}, and {\it dashed-dotted blue lines} show our central value, statistical and statistical plus systematic error bands, respectively, as determined from these fits. See text for details.}
\label{fig:massrange_NLO}
\end{minipage}
\end{center}
%
\end{figure}

For the chiral fits, we first applied several different cuts for the heaviest mass included in the fit range. These cuts are indicated by the horizontal dashed lines in the landscape plot of our simulated points, Fig.~\ref{fig:landscape}. In a next step, we studied the effects of excluding one or more lattice spacings from the fitted data. In the end, judging by the resulting (uncorrelated) $\chi^2/{\rm d.o.f}$ of the fit and reaching a plateau for the fitted parameters, we chose $135\,{\rm MeV}\,\leq\, M_\pi\,\leq\,240\,{\rm MeV}$ and $1/a\,\geq\,1.6\,{\rm GeV}$ to be our preferred fit range. This choice is indicated by the solid blue lines in Fig.~\ref{fig:landscape}. The resulting combined global fit to the unconstrained fit formulae is shown in Fig.~\ref{fig:fitNLO_135_240_fine}. In Figure \ref{fig:massrange_NLO} we show the impact of the different mass cuts on the fitted parameters and the resulting (uncorrelated) $\chi^2/{\rm d.o.f.}$ using only ensembles with $1/a\,\geq\,1.6\,{\rm GeV}$. For the systematic uncertainty we decided to take the variation of all these points with respect to the central value from our preferred fit. The central value and statistical uncertainty from our preferred fit is indicated in each panel by a solid and dashed line, respectively, whereas the total (statistical plus systematic) uncertainty is indicated by a dashed-dotted line. Details on the impact of removing one or more lattice spacing from the fits can be found in \cite{Borsanyi:2012zv}.

\vspace*{\afterFigure}

\begin{table}[b!]
\begin{center}
\begin{tabular}{lrrr}
\hline\hline
 & \multicolumn{1}{c}{unconstrained} & \multicolumn{1}{c}{parameter-reduced} & \multicolumn{1}{c}{final} \\\hline
$\chi^{\rm phys}/(10^{-2} {\rm GeV}^2)$ & $1.8578(17)(39)$  &  $1.8639(18)(44)$  & $1.8609(18)(74)$ \\
$f/{\rm MeV}$                   &  $122.70(08)(41)$  &  $122.73(06)(28)$ & $122.72(07)(35)$ \\
$\Lambda_3/{\rm MeV}$           &  $628(23)(57)$    & $678(40)(119)$ & $653(32)(101)$ \\
$\Lambda_4/{\rm MeV}$           &  $1,012(16)(83)$   & $1,006(15)(71)$ & $1,009(16)(77)$\\
\hline\hline
\end{tabular}
\end{center}
\vspace*{\closercaption}
\caption{Results with statistical and systematic uncertainties for fitted parameters from unconstrained {\it (first column)} and parameter-reduced {\it (second column)} NLO SU(2) ChPT fits. Note that in the parameter-reduced case only the first two parameters are actual free fit parameters, while the other two are derived therefrom. The {\it third column} shows our combined final results, see text for details.} 
\label{tab:SU2results}
\end{table}
\vspace*{\afterTable}

We repeated the same analysis with the parameter-reduced SU(2) ChPT-formulae, which have been constrained to hit the physical point. In Table \ref{tab:SU2results} we summarize the obtained fit parameters from each ansatz. Note that for the parameter-reduced fit only the first two lines are fitted parameters while the two low-energy scales $\Lambda_3$, $\Lambda_4$ are derived from these. The table also contains our final estimate for the NLO SU(2) LECs, which were obtained by combining the results from the two fit ans\"atze in the following way: we averaged the central values and the statistical uncertainties. For the square of the systematic uncertainty we sum the squares of the average systematic uncertainty and the spread of the central values.

\begin{figure}[t!]
\begin{center}
\includegraphics*[width=.4\textwidth]{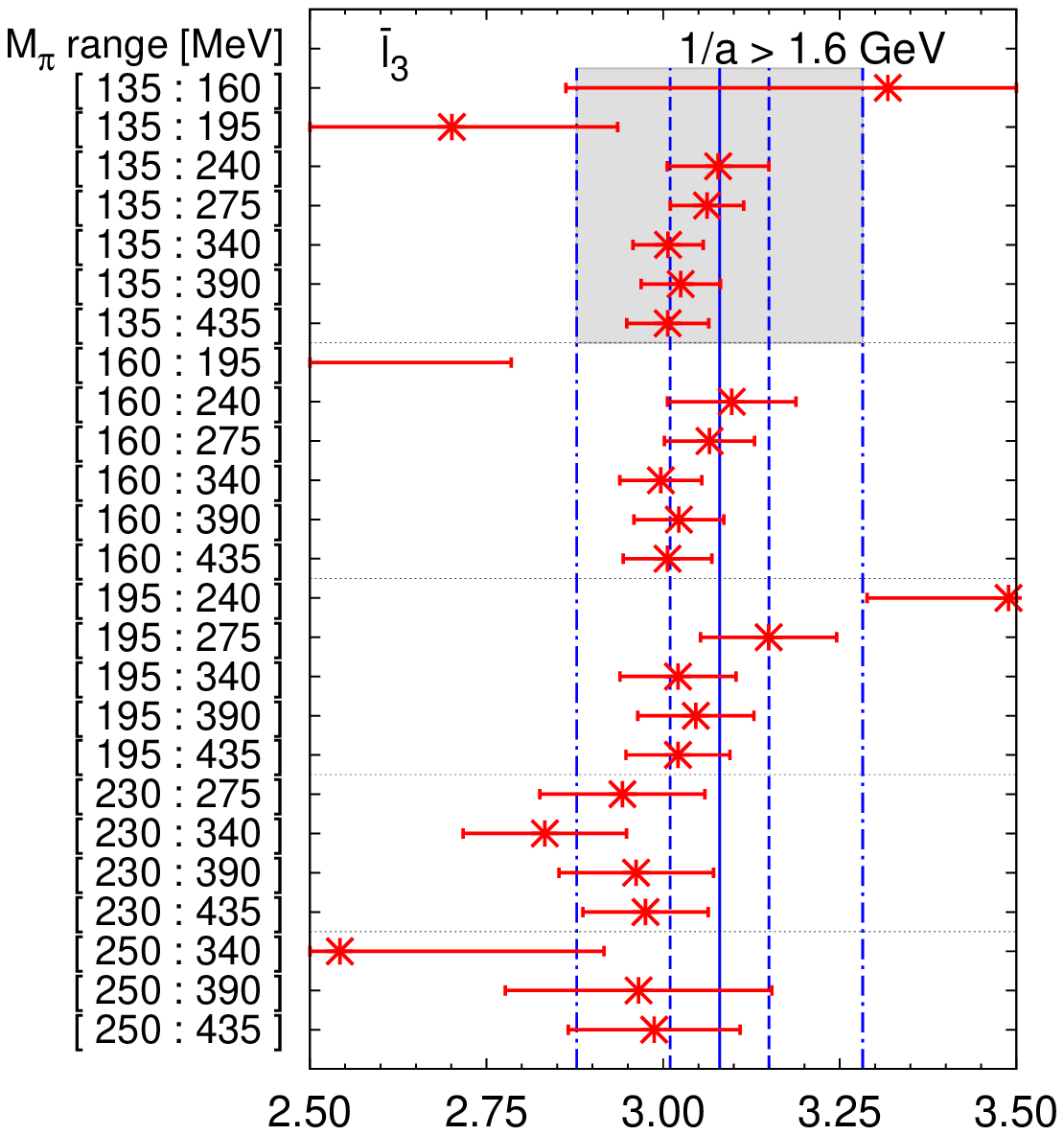}
\includegraphics*[width=.4\textwidth]{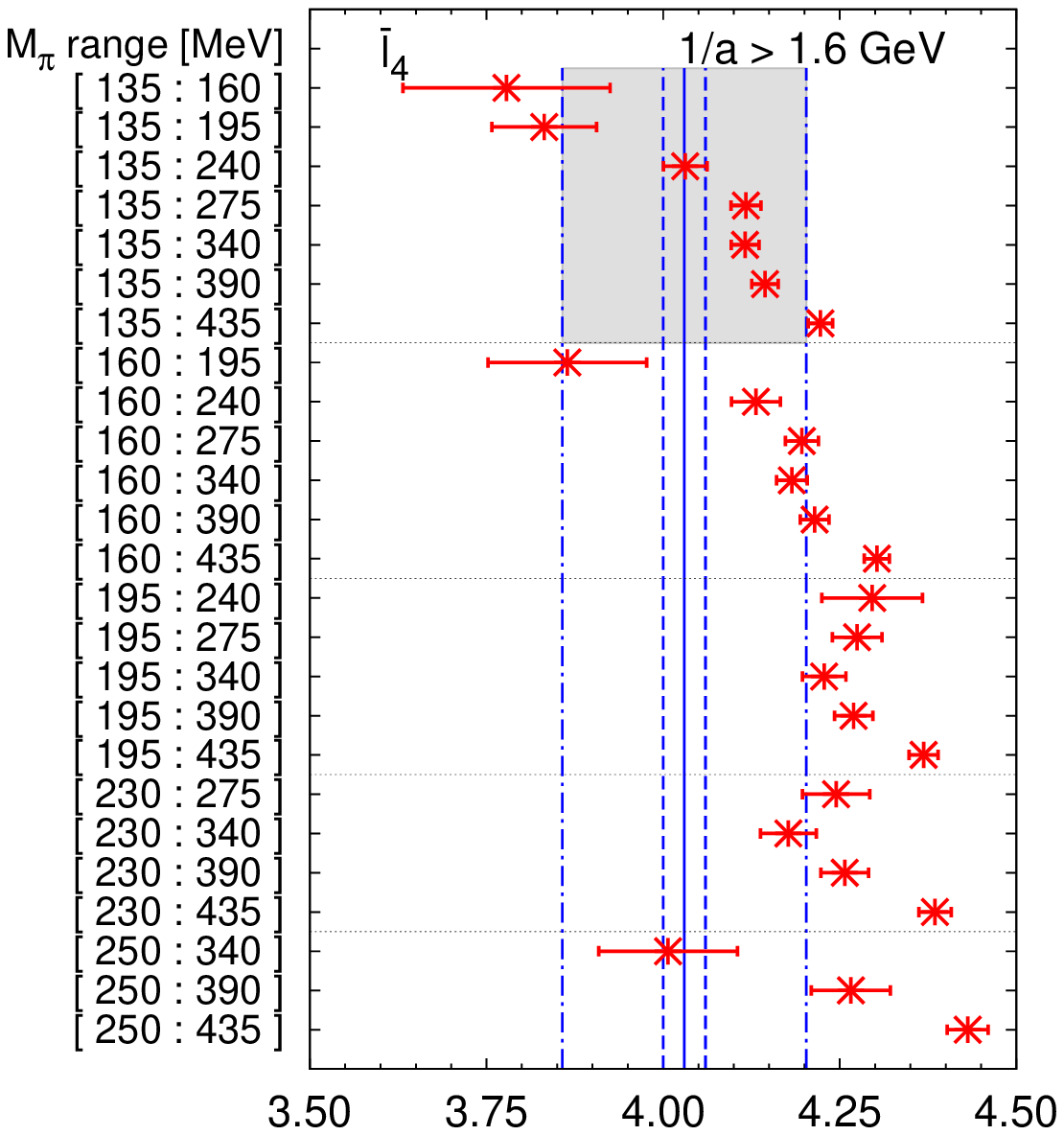}
\end{center}
\vspace*{\closercaption}
\caption{Results for the LECs $\bar{\ell}_3$ {\it(left panel)} and $\bar{\ell}_4$ {\it(right panel)} from unconstrained NLO SU(2) ChPT fits for different mass ranges including (above top-most dashed horizontal line) and excluding near-physical masses. Only ensembles with $1/a\geq1.6\,{\rm GeV}$ were included in the fits. {\it Blue lines} indicate our central value and error bands in this set-up.}
\label{fig:SU2_nophys}
\end{figure}
\vspace*{\afterFigure}

\begin{figure}[t!]
\begin{center}
\includegraphics*[width=.4\textwidth]{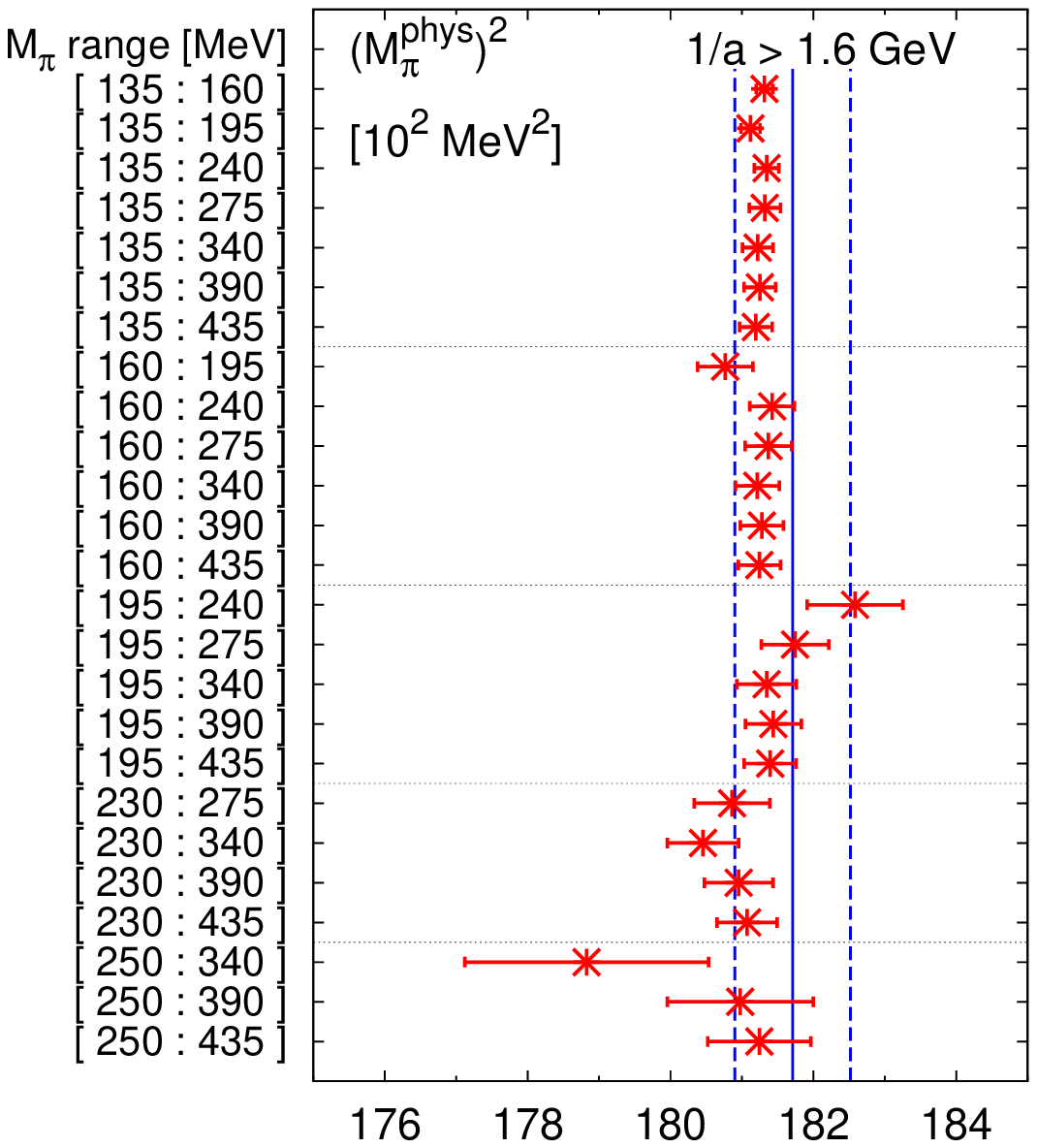}
\includegraphics*[width=.4\textwidth]{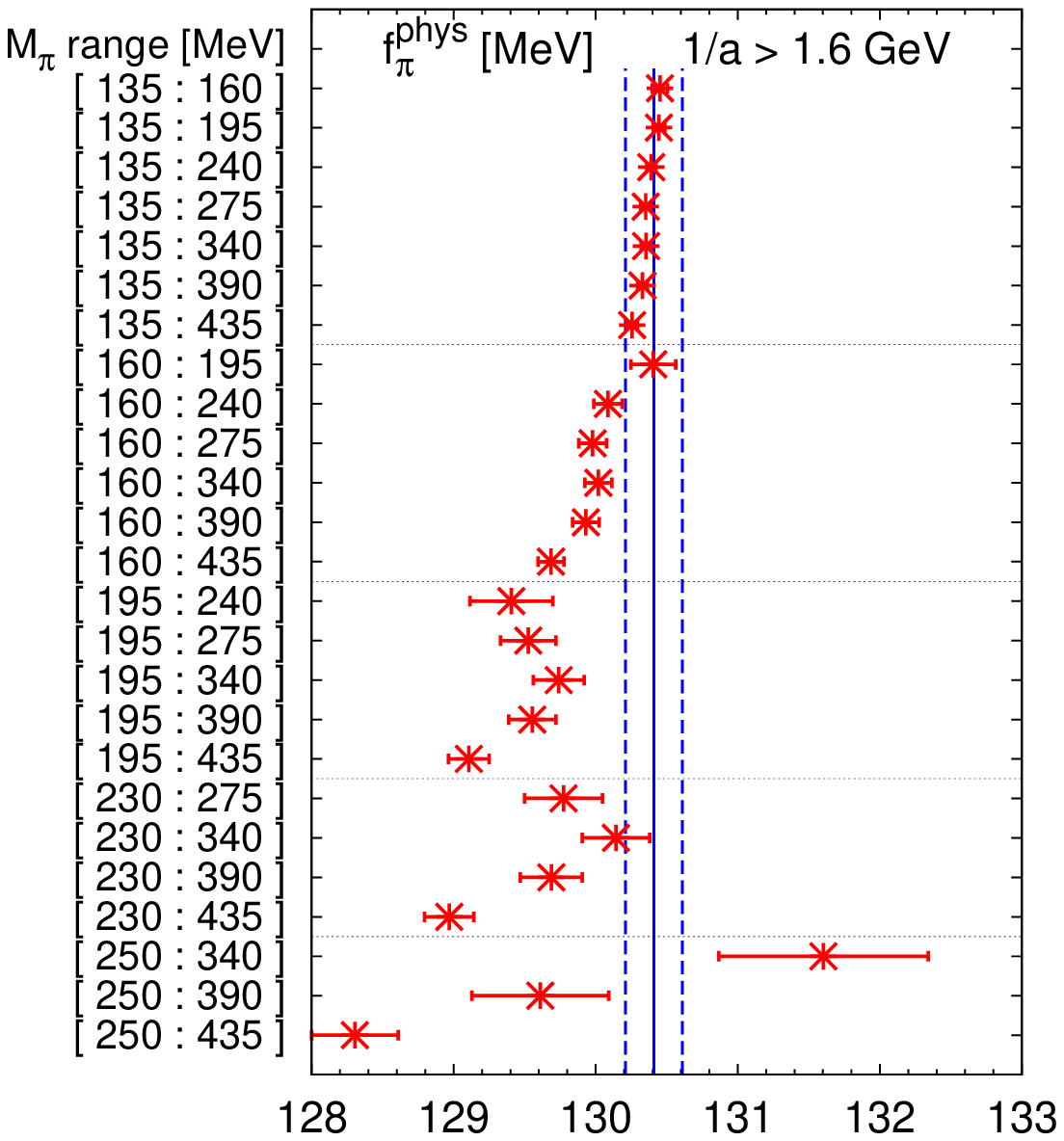}
\end{center}
\vspace*{\closercaption}
\caption{Extrapolated physical pion mass squared {\it(left panel)} and decay constant {(\it right panel)} from unconstrained NLO SU(2) ChPT for different mass ranges including (above top-most dashed horizontal line) and excluding near-physical masses. {\it Blue lines} indicate the experimental values and error bands from \cite{Colangelo:2010et,Nakamura:2010zzi}.}
\label{fig:SU2_extrap}
\end{figure}
\vspace*{\afterFigure}

In addition to our main NLO SU(2) ChPT fits, we also varied the lower cut on the pion masses in the unconstrained fits. Especially, this is of interest because nowadays many lattice simulations still do not include the physical point and use ChPT to extrapolate towards it. In Figure \ref{fig:SU2_nophys} we show how, e.g., the low-energy constants $\bar{\ell}_3$ and $\bar{\ell}_4$ (related to the scales $\Lambda_3$ and $\Lambda_4$, respectively) change due to the various lower and higher mass cuts (in the upper part of each plot (grey shaded area) the results of fits including the physical point are shown, which have been used to obtain the systematic uncertainty, see above). One can see that $\bar{\ell}_3$, which predominantly influences the quark mass dependence of $M_\pi^2$, is still in agreement with results from fits including the near physical points, whereas $\bar{\ell}_4$, which predominantly influences the quark mass dependence of $f_\pi$, shows some deviations when the physical point is excluded. %
In Figure \ref{fig:SU2_extrap} we compare the pion mass and decay constant extrapolated from our unconstrained NLO SU(2) ChPT fits (including and excluding the near physical points) to the experimental values \cite{Colangelo:2010et,Nakamura:2010zzi}. Also here the pion decay constant shows more deviations from the expected result, once more and more lighter masses are excluded from the fits. In our opinion, these observations illustrate the danger inherent in applying NLO SU(2) ChPT-formulae to lattice data lacking data points with light enough pion masses.

\section{Fits to NNLO SU(2) ChPT}
\label{subsec:fit.nnlo}
\vspace*{\closersection}

\begin{figure}[t!]
\begin{center}
\includegraphics*[width=.45\textwidth]{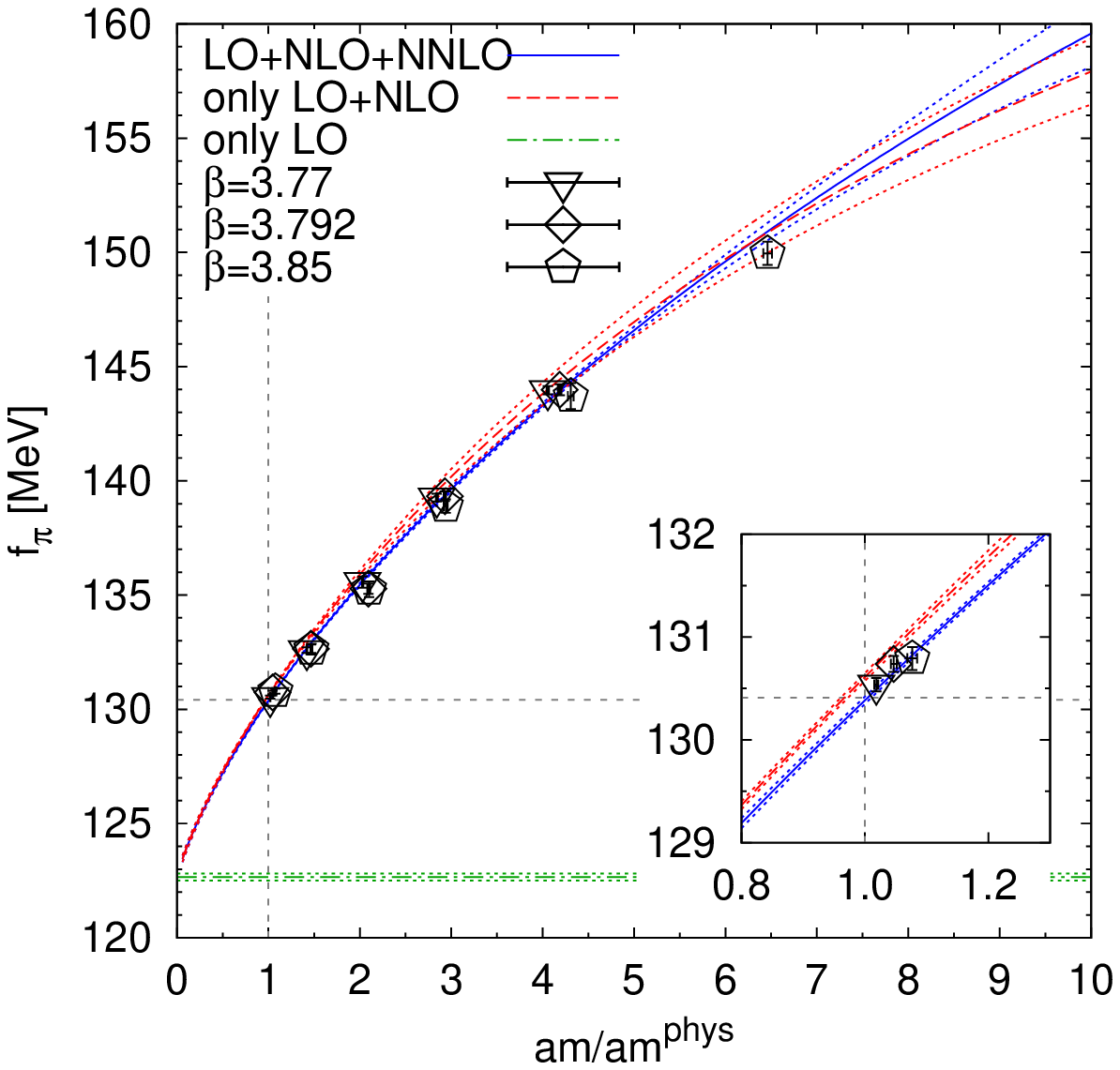}%
\includegraphics*[width=.45\textwidth]{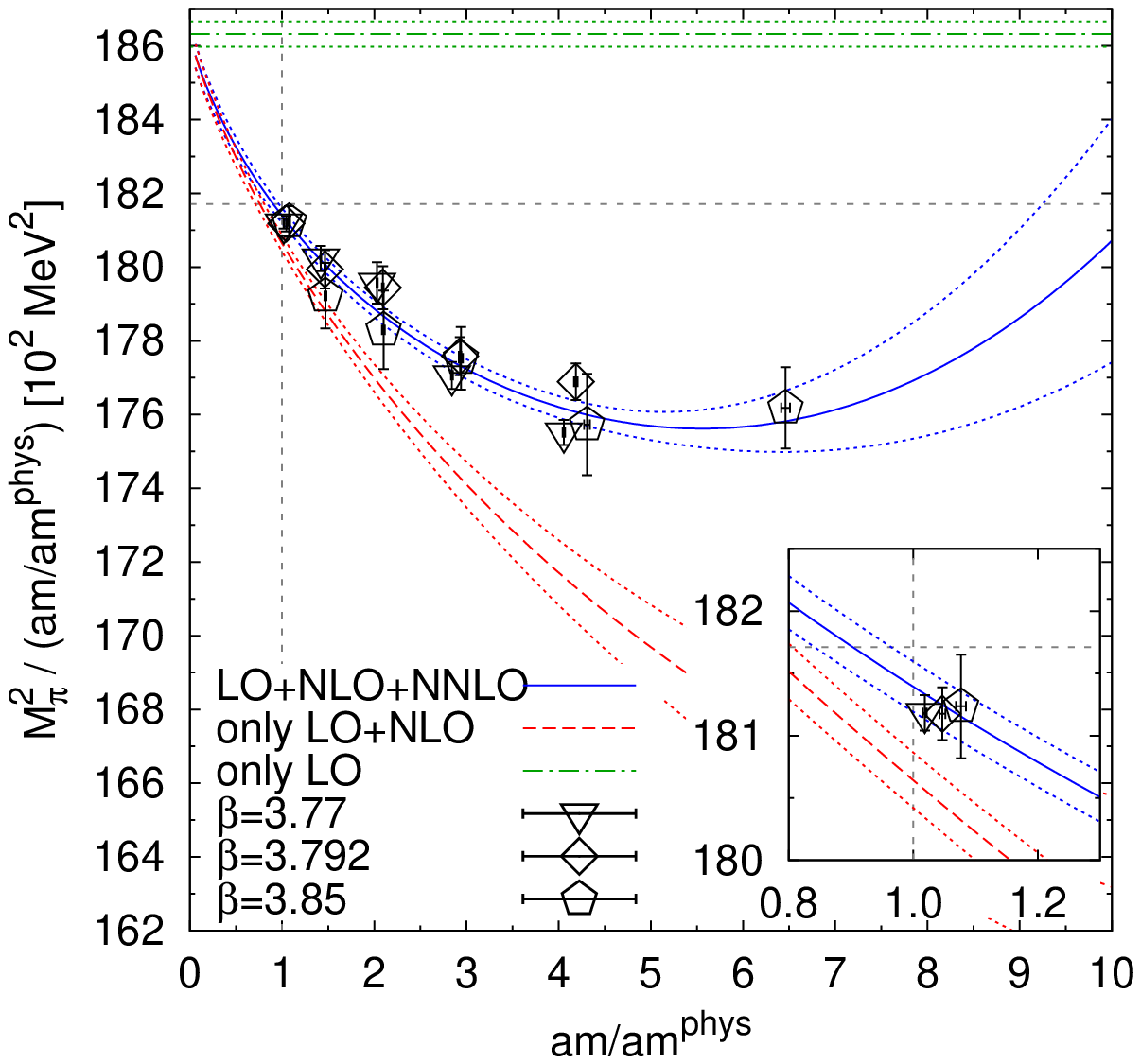}
\end{center}
\vspace*{\closercaption}
\caption{Combined global fit to NNLO SU(2) ChPT using priors for $\Lambda_{12}$, $k_{M^2}$ and $k_f$. Only ensembles with $1/a\geq 1.6\,{\rm GeV}$ and $135\,{\rm MeV}\leq M_\pi\leq340\,{\rm MeV}$ are included in the fit (only data points included in the fit range are shown in these plots).}
\label{fig:fit_NNLO}
\end{figure}
\vspace*{\afterFigure}

\begin{figure}[t!]
\begin{center}
\includegraphics*[width=.4\textwidth]{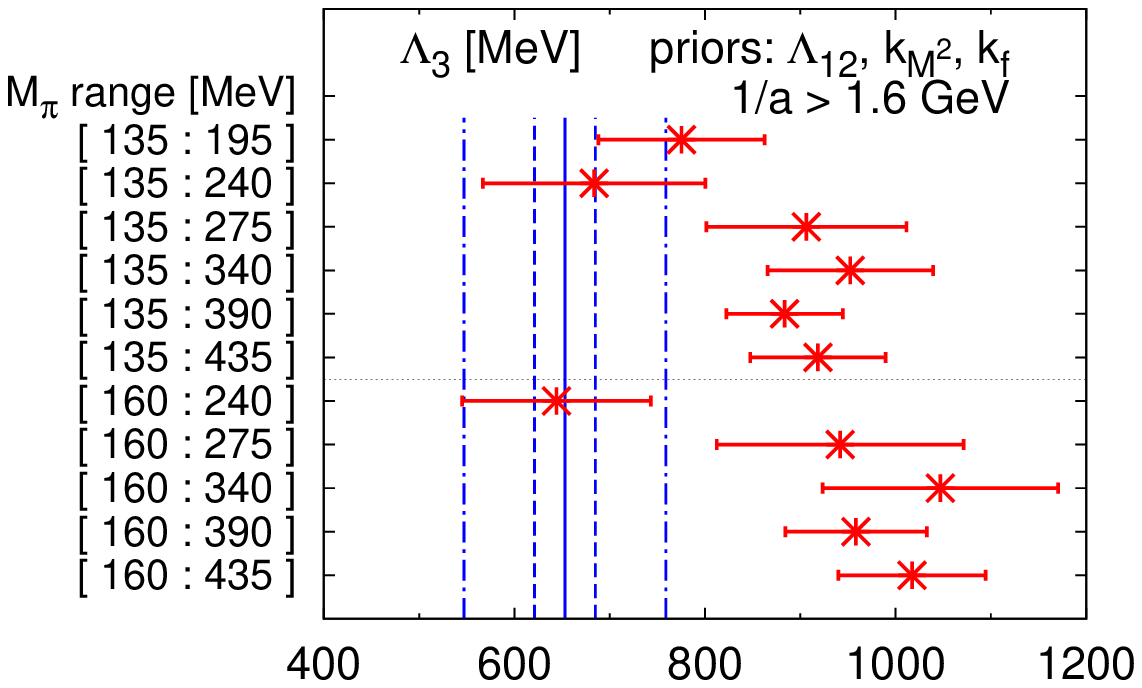}%
\includegraphics*[width=.4\textwidth]{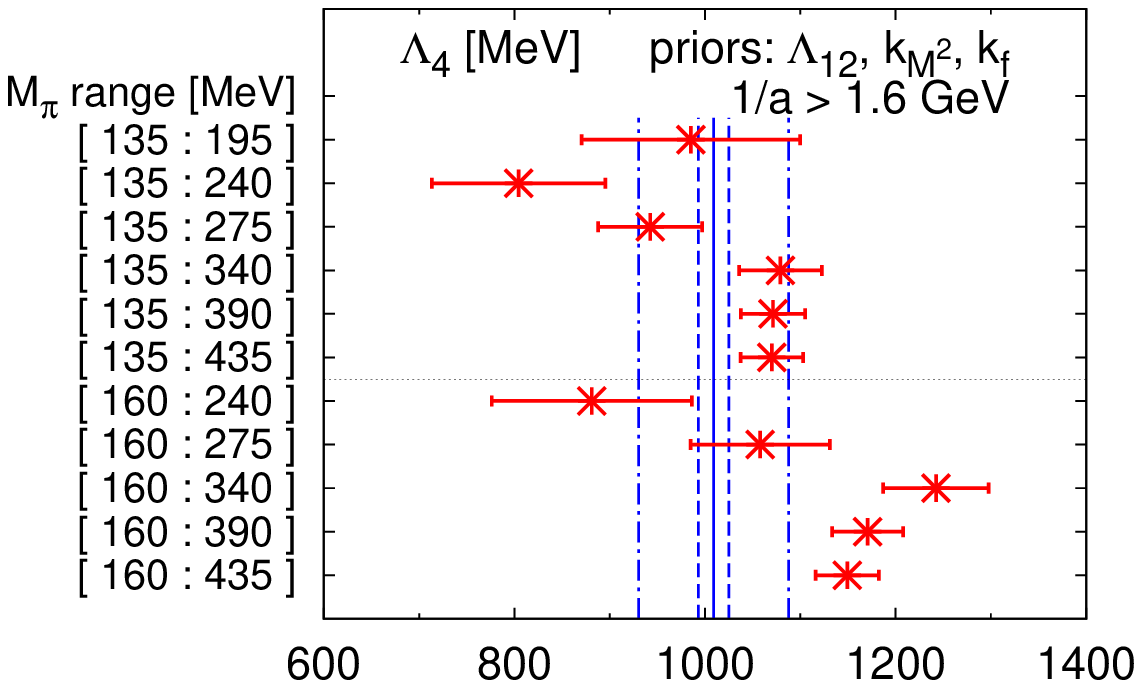}
\end{center}
\vspace*{\closercaption}
\caption{Results for $\Lambda_3$ {\it (left panel)} and $\Lambda_4$ {\it (right panel)} from NNLO SU(2) ChPT fits to different mass ranges using priors for $\Lambda_{12}$, $k_{M^2}$, and $k_f$. Only ensembles with $1/a\geq 1.6\,{\rm GeV}$ were included in the fits. {\it Blue lines} indicate our final central value and error bands.}
\label{fig:results_NNLO}
\end{figure}
\vspace*{\afterFigure}

To check the influence of higher orders in SU(2) ChPT on our results for the LECs, we also extended our fits to include the next-to-next-to-leading order (NNLO) contributions. In our set-up for the fits of the meson masses and decay constants, we had to add three new fit parameters: the low-energy scale $\Lambda_{12}$, which we defined as a combination of the NLO low-energy scales $\Lambda_1$ and $\Lambda_2$: $\log\Lambda_{12}^2=(7\log\Lambda_1^2+8\log\Lambda_2^2)/15$ and two parameters for the NNLO-contributions in the meson mass and decay constant dependence, $k_{M^2}$ and $k_f$, respectively (see \cite{Borsanyi:2012zv} for details). Unfortunately, our amount of data at fine enough lattice spacings and light quark masses did not turn out to lead to stable fits. Therefore, in the end we opted for using priors on the three additional fit parameters in the NNLO-fits. Details on the choice for the priors are reported in \cite{Borsanyi:2012zv}. In Figure \ref{fig:fit_NNLO} we show a typical fit of our data to the NNLO formulae using priors. Since we were not able to resolve the new parameters in a satisfactory way without a priori input we refrain from quoting results for these parameters. But we would like to stress that the change in the determined values for the NLO-LECs is only minor as can be seen from Fig.~\ref{fig:results_NNLO} where we show the results for $\Lambda_3$ and $\Lambda_4$ from a NNLO-fit (using priors for $k_{M^2}$, $k_f$, and $\Lambda_{12}$) at different mass ranges. There the vertical lines show our final results and error bands from our NLO-fits.

\section{Conclusions}
\vspace*{\closersection}

We determined the following SU(2) LECs from our NLO ChPT fits to meson masses and decay constants measured on staggered 2+1 flavor lattice simulations of QCD, cf.\ also the right-most column of Tab.~\ref{tab:SU2results}.
\begin{eqnarray*}
 2Bm^{\rm phys}\;=\;1.8609(18)_{\rm stat}(74)_{\rm syst}\,\cdot\,10^{-2}\,{\rm GeV}^2\,, && f\;=\;122.72(07)_{\rm stat}(35)_{\rm syst}\,{\rm MeV}\,,\\
 \bar{\ell}_3\;=\;3.16(10)_{\rm stat}(29)_{\rm syst}\,, && \bar{\ell}_4\;=\;4.03(03)_{\rm stat}(16)_{\rm syst}\,.
\end{eqnarray*}
(Here we quoted the LECs $\bar{\ell}_3$ and $\bar{\ell}_4$, which are related to the scales $\Lambda_3$ and $\Lambda_4$, respectively.) In addition, we also provide the ratio of the extrapolated physical decay constant to its value in the chiral limit $f^{\rm phys}_\pi/f\;=\;1.0627(06)_{\rm stat}(27)_{\rm syst}$. Using the value determined for the average light quark mass in \cite{Durr:2010vn,Durr:2010aw}, we obtain from our fitted value for $\chi^{\rm phys}$ the condensate parameter
\[ B^{\overline{\rm MS}, 2{\rm GeV}}\;=\;2.682(36)_{\rm stat}(39)_{\rm syst}\,{\rm GeV}\,,\;\;\;\Sigma^{\overline{\rm MS}, 2{\rm GeV}}\;=\;\big(272.3(1.2)_{\rm stat}(1.4)_{\rm syst}\,{\rm MeV}\big)^3\,, \]
where the condensate was obtained by multiplying $B$ with our result for $f^2/2$.

The speaker acknowledges support from the DFG SFB/TR 55 and the EU grants PIRG07-GA-2010-268367 and PITN-GA-2009-238353 (ITN STRONGnet).


\vspace*{-.4cm}
\bibliography{references}
\bibliographystyle{h-physrev5}

\end{document}